# X-ray studies: Phase transformations and microstructure changes

*C. Scheuerlein, M. Di Michiel*

## G1.1.2.1 Introduction

Conventional materials characterisation of superconducting wires or tapes implies the destructive preparation of the samples by cutting, grinding and polishing, for instance for microscopic studies. A destructive sample preparation is also required for X-ray diffraction (XRD) experiments with laboratory diffractometers, which commonly use Cu Kα radiation ($E_{Cu-K\alpha} \sim 8.03$ keV) with a penetration depth of some tens of μm in metallic samples.

In contrast, both neutrons and high energy photons can penetrate mm-thick strongly absorbing superconductors, which enables non-destructive diffraction experiments, for instance with Bi-2223 tapes [i,ii]. The X-ray transmission through typical Ta alloyed $Nb_3Sn$ wires as a function of the X-ray energy is presented in Figure G1.1.2.1. X-ray energies above 50 keV enable non-destructive experiments in transmission geometry with $Nb_3Sn$ wires [iii].

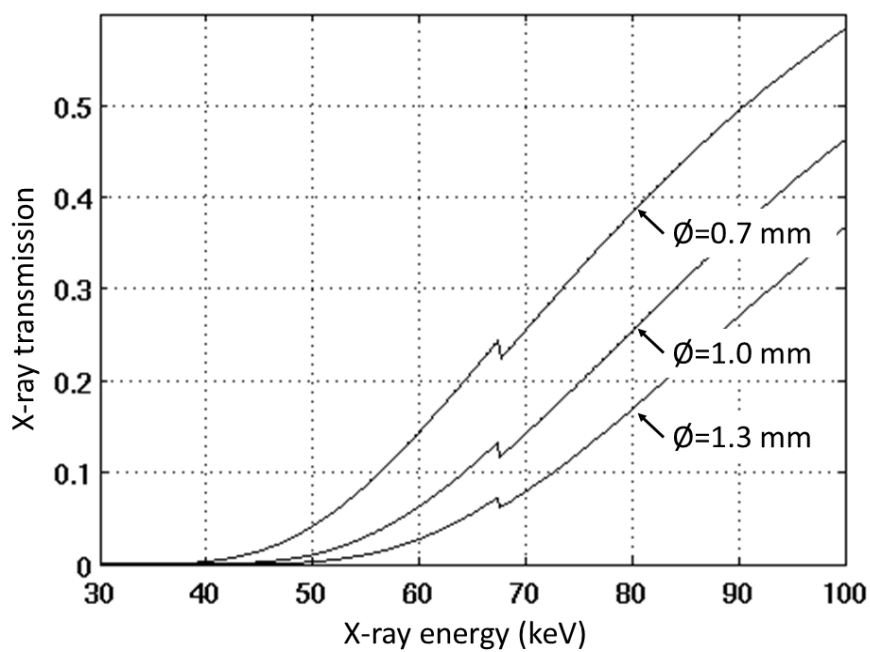

**Figure G1.1.2.1.** X-ray transmission through Ta alloyed $Nb_3Sn$ wires of equal composition with different diameters.

Typical acquisition times of neutron diffraction pattern of individual superconducting wires are in the order of hours [iv]. State-of-the-art high energy synchrotron sources can provide very high monochromatic photon flux densities, and diffraction measurements can be performed within seconds when using fast read-out area detectors in Debye-Scherrer transmission geometry. This chapter focuses on high energy synchrotron radiation *in situ* studies of entire processes, which can be much faster than conventional materials studies that require a series of samples to be prepared after different processing steps. The non-destructive *in situ* studies also avoid experimental uncertainties caused by sample quenching, sample inhomogeneities and preparation artefacts.

The relatively small scattering angles of high energy X-rays, facilitate to add auxiliary equipment, for instance a furnace for heat treatment (HT) studies. For the experiments presented here two furnaces of the ESRF ID15 beamline have been used. For *in situ* studies in inert gas or air at ambient pressure the ID15 diffraction and tomography furnace was used (Figure G1.1.2.2.a). After alignment with respect to the X-ray beam, the position of this furnace remains fixed during the experiment. The sample is mounted onto a ceramic stick that enters the furnace from a bottom hole. For sample alignment and rotation the ceramic stick is mounted onto a goniometer and the rotation and translation stages. The thin Al foil windows at both sides of the furnace are nearly transparent for high energy photons. The furnace temperature can be regulated using the temperature reading of a thermocouple that can be inserted into the furnace through the bottom hole. Large sample temperature uncertainties can be avoided when the thermocouple is spot welded onto the sample.

For overpressure *in situ* studies a set-up consisting of a capillary furnace around a high pressure cell made of a single crystal sapphire tube has been used (Figure G1.1.2.2.b). This furnace was developed for combined *in situ* high energy X-ray diffraction and mass spectrometry investigations during catalysed gas/solid or liquid/solid reactions [v]. The connection of the high pressure cell to the pressure controller is done with a flexible stainless steel line that allows rotating the high pressure cell up to 360° during the acquisition of diffraction patterns. For the study of superconducting wires the high pressure cell was modified such that a thermocouple can be spot welded onto the superconductor sample [vi].

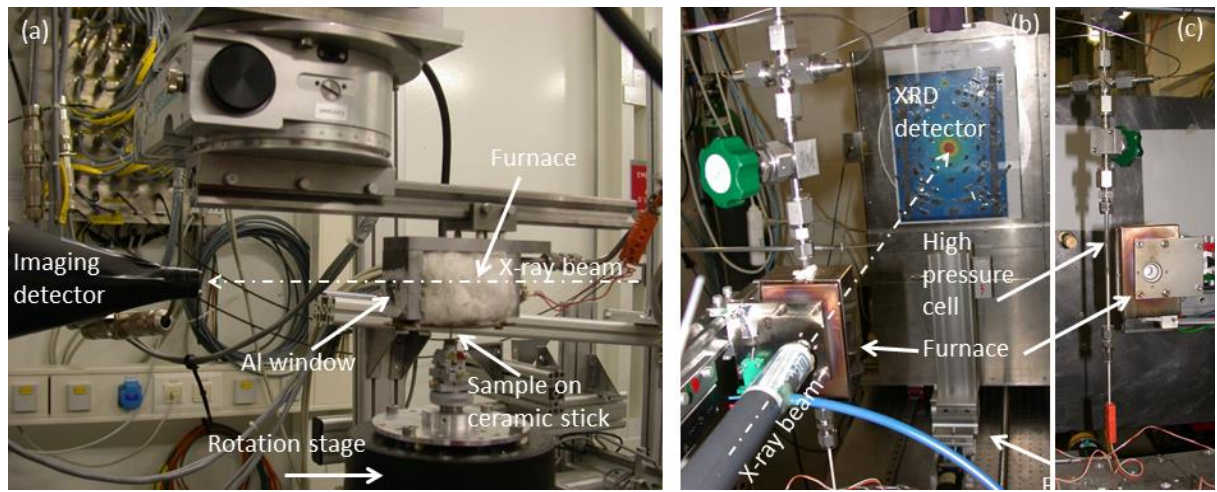

**Figure G1.1.2.2.** (a) ID15 furnace for *in situ* XRD and tomography at ambient pressure. (b) Furnace for *in situ* XRD at pressures up to 200 bar in measurement position and (c) capillary furnace withdrawn from the high pressure cell.

Other sample environments are possible too, for instance a cryostat and a tensile rig can be added. This makes it possible to study the superconductor electromechanical behaviour and the damage development by XRD measurements at well-defined uniaxial tensile stress or strain at cryogenic temperatures, for instance in liquid Helium [iii,vii,viii] or in liquid Nitrogen [ix]. Lattice distortions, superconducting properties and mechanical properties of high temperature superconductors can be measured simultaneously [ix].

X-ray absorption micro-tomography (μ-CT) can provide three dimensional images of the superconductor bulk. A spatial resolution of μ-CT in the order of 1 μm is today routinely obtainable with both laboratory and synchrotron sources. Fast μ-CT [x,xi], where the acquisition of the about 1000 radiographs needed for the reconstruction of one tomogram lasts not more than 1 minute, is required for *in situ* studies of microstructural changes and porosity formation during the processing of superconductors with temperature ramp rates in the order of 100 °C/h.

Different synchrotron techniques, for instance high energy XRD and μ-CT, can be combined in one experiment. This combination has been pioneered at the ESRF ID15A beamline for an *in situ* study of the void growth mechanisms in Nb₃Sn wires [xii]. This was achieved using two X-ray beams, a high intensity filtered white X-ray beam for μ-CT and a monochromatic X-ray beam (energy 88 keV, energy bandwidth 0.1 keV) for the XRD measurements [xiii]. The sample and the furnace needed to be aligned in both X-ray beams, and during the entire HT cycle, they were continuously moved from the white beam to the monochromatic beam for the alternating XRD and μ-CT measurements. After the installation of the new ID15 insertion device in 2008 the flux density of high energy monochromatic photons has been further increased such that XRD and fast μ-CT can now be performed both with the same monochromatic photon beam.

The goal of this chapter is to illustrate the potential of high energy synchrotron radiation experiments for *in situ* studies of the processing of superconductors. We present case studies describing the Nb₃Sn wire diffusion HT, the transformation HT of Nb₃Al precursor wires, and the melt processing HT of Bi-2212 wires.

## G1.1.2.2 Nb₃Sn diffusion HT

### *G1.1.2.2.1 Phase transformations during the diffusion HT of Nb₃Sn superconductors*

The Nb₃Sn phase in multifilament wires is produced during a diffusion HT where the precursor elements Nb and Sn interdiffuse with the Cu matrix, forming various intermetallic phases and finally the superconducting Nb₃Sn [xiv]. The intermediate phase transformations can degrade the microstructural and microchemical homogeneity of the fully reacted superconductor. This is most easily observed in the tubular strand types (Powder-in-Tube (PIT) [xv] and Tube Type [xvi]), where typically 25% of the Nb₃Sn volume consists of coarse grains that are not well connected and cannot conduct significant supercurrents.

High energy synchrotron X-ray diffraction is an excellent tool to monitor the phase changes in superconducting wires *in situ* during the processing HT [xvii,xviii]. As an example, Sn, Cu₆Sn₅, NbSn₂, Nb₆Sn₅, Nb₃Sn, and a CuNbSn ternary phase can be identified in the diffraction pattern that have been acquired during Nb₃Sn PIT wire HT (Figure G1.1.2.3). The respective temperature intervals where these phases are present are easily revealed in the sequence of diffractograms.

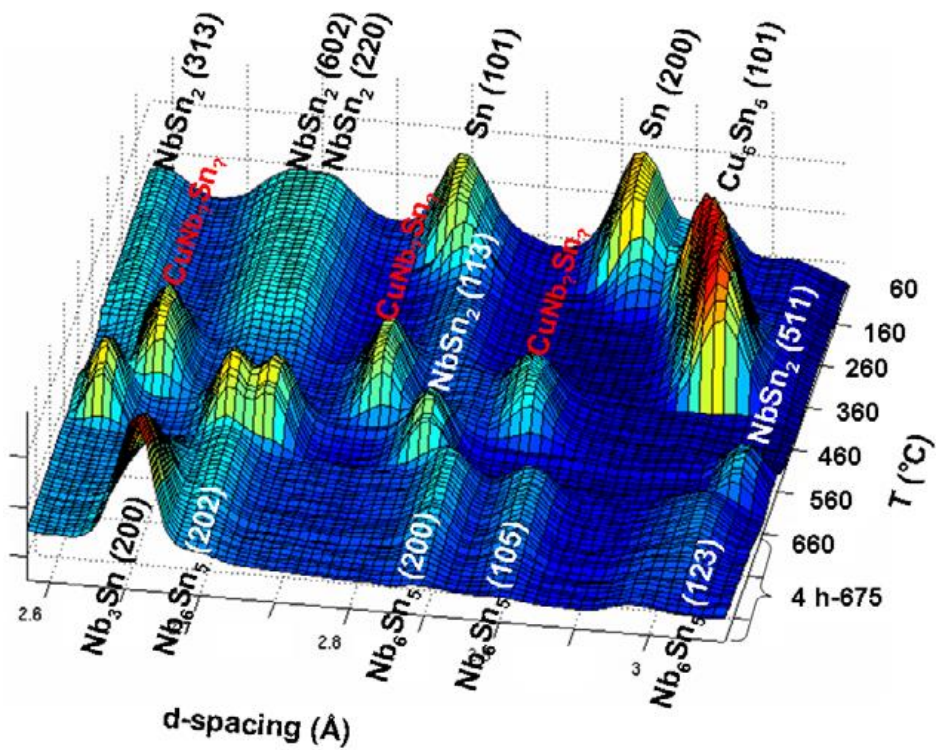

**Figure G1.1.2.3.** Summary of the diffraction patterns acquired *in situ* during the Nb₃Sn PIT wire reaction HT. © IOP Publishing. Reproduced with permission. All rights reserved.

*In situ* XRD measurements during the reaction HT of Restacked Rod Process (RRP) type [xix] and Tube Type [xx] Nb₃Sn wires revealed a similar phase sequence. In particular CuNbSn, NbSn₂ and Nb₆Sn₅ are detected in the high $J_c$ strands. In the RRP type wire the amount of these phases is comparatively small, which presumably explains the relatively small volume fraction of Nb₃Sn coarse grains in the fully processed RRP wire. During the processing HT of a low Sn content Internal Tin wire [xxi] a markedly different phase sequence is observed, and in particular the phases CuNbSn, NbSn₂ and Nb₆Sn₅ are not formed [xii] (see G1.1.2.5).

### G1.1.2.2.2 Nb₃Sn nucleation and growth

Nb₃Sn nucleation and growth in multifilament wires is accompanied by changes of the Sn content distribution, and the Nb₃Sn grain size distribution. These microstructure and composition changes, which have a strong influence on $J_c$ [xxii,xxiii], can be followed *in situ* by high energy synchrotron XRD measurements.

By monitoring the Nb₃Sn diffraction peak area evolution the Nb₃Sn formation kinetics in different wires can be compared. In an Internal Tin wire with low Sn content the Nb₃Sn phase growth follows a parabolic law [xvii], indicating that in this wire Nb₃Sn growth is diffusion controlled. This is in contrast to the Nb₃Sn phase growth in state-of-the-art high Sn content RRP and PIT type wires, where Nb₃Sn growth is not purely diffusion controlled [xvii,xix].

Figure G1.1.2.4 compares the Nb₃Sn growth in a RRP wire with 80 µm subelement size with that in PIT wires with 30 µm and 50 µm subelement size. All wires followed the same HT cycle with 100 °C/h heating rate and three isothermal steps (4 h-700 °C, 1 h-800 °C and 1 h-900 °C). Usually the processing peak temperature of Nb₃Sn superconductors does not exceed 700 °C. Here the 800 °C and 900 °C plateaus were added to explore the full reaction within a duration that allows to perform *in situ* synchrotron experiments. It is assumed that in the three wires the maximum possible amount of $Nb_3Sn$ was formed during the HT. Therefore, the maximum $Nb_3Sn$ peak areas measured during the HT cycles can be normalised, and the $Nb_3Sn$ growth kinetics in the wires with different elemental composition and architecture can be compared.

In the RRP wire with 80 µm subelement size $Nb_3Sn$ is already detected at about 540 °C and about 80% of the maximum possible $Nb_3Sn$ volume is formed after the 4 h-700 °C plateau. In the PIT wires $Nb_3Sn$ is first detected during the 700 °C plateau, and the $Nb_3Sn$ formation kinetics and the duration needed to transform $Nb_6Sn_5$ entirely into $Nb_3Sn$ are significantly influenced by the subelement size.

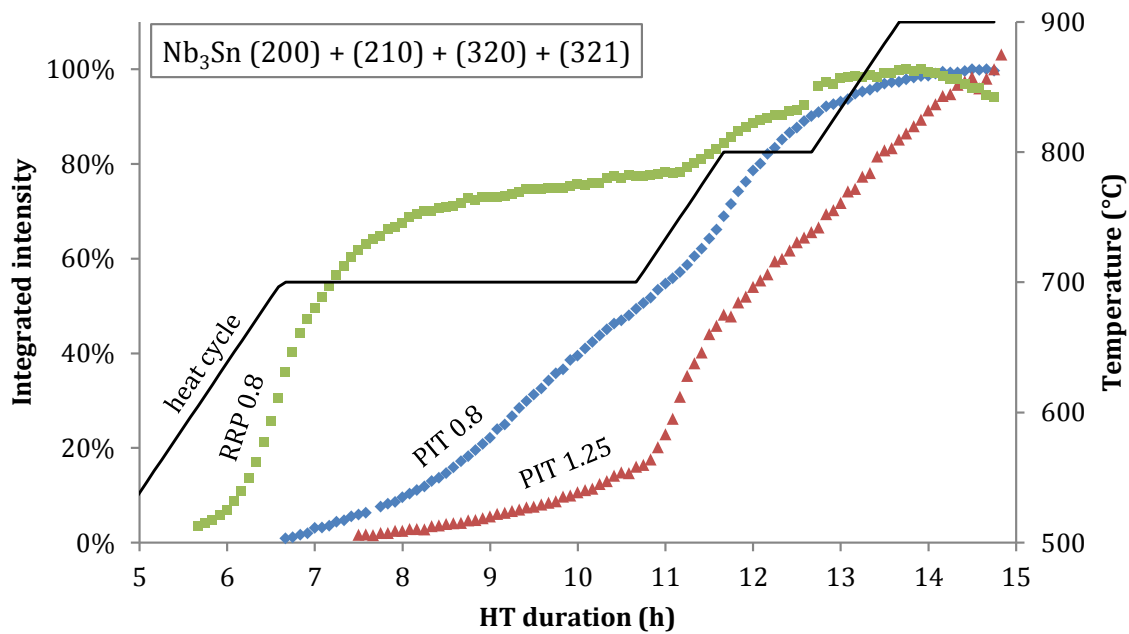

**Figure G1.1.2.4.** Evolution of the integrated intensity of prominent $Nb_3Sn$ peaks in the PIT Ø=0.8 mm, PIT Ø=1.25 mm and RRP Ø=0.8 mm wires during identical HT. Courtesy J. Kadar.

Keeping a small grain size for flux pinning and at the same time have maximum $Nb_3Sn$ volume and Sn content are conflicting needs of the HT. In ideally homogeneous wires, the $Nb_3Sn$ grain size and Sn content evolution can be monitored simultaneously with the $Nb_3Sn$ volume by XRD measurements. The Sn content can be determined from $Nb_3Sn$ lattice parameter measurements [xxiv]. Assuming that the $Nb_3Sn$ grains nucleate and grow in a nearly stress-free state, the decrease of diffraction peak width after deconvolution of the instrument function is associated to the increase of grain size, and the use of the Scherrer formula allows for a rough calculation of the mean crystallite size [xxv]. In order to monitor crystallite sizes up to 200 nm the diffraction experiment needs to be optimised for minimising instrumental peak broadening.

Since the RRP wire has a comparatively homogeneous $Nb_3Sn$ microstructure it has been selected for the $Nb_3Sn$ nucleation and growth *in situ* study [xxvi]. Figure G1.1.2.5 compares the changes of the $Nb_3Sn$ volume with the average crystallite size evolution (a) and the average Sn content (b) during 100 °C/h HT with the three isothermal steps at 700 °C, 800 °C and 900 °C.

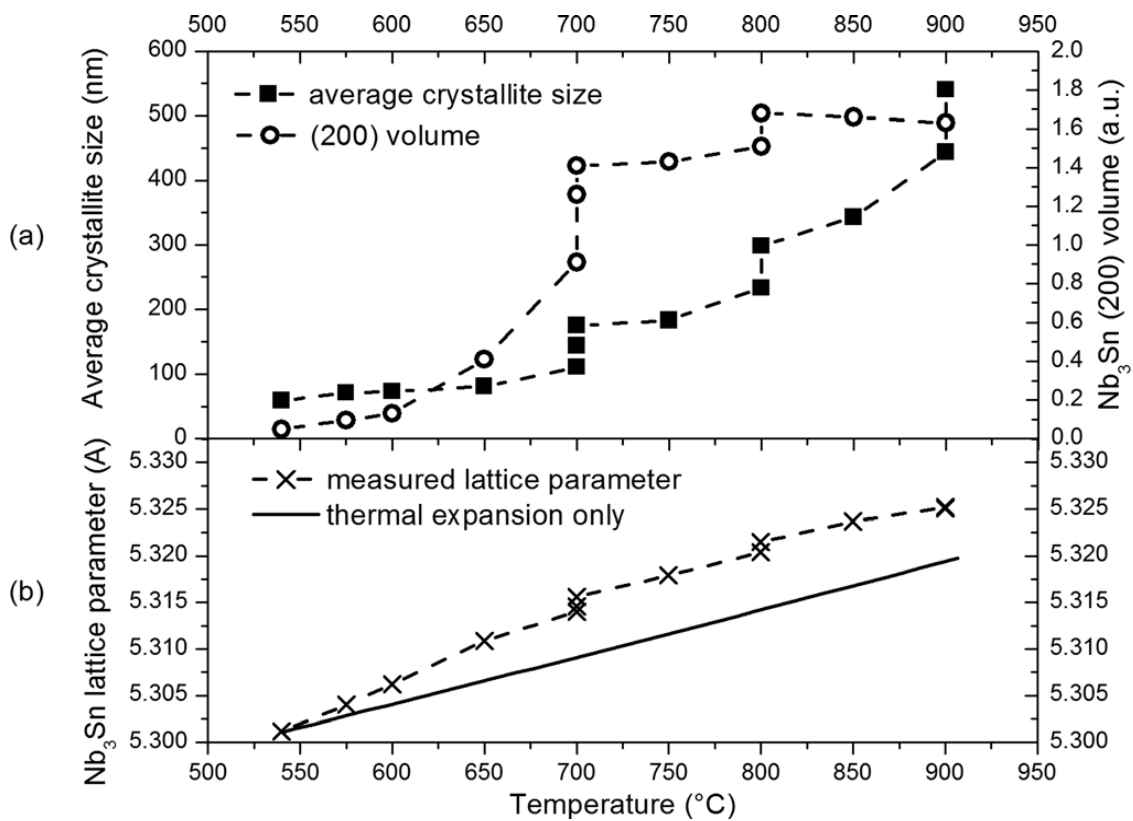

**Figure G1.1.2.5.** Average Nb₃Sn crystallite size and volume from Nb₃Sn (200) reflection and (b) Nb₃Sn lattice parameter as a function of temperature. The relative lattice parameter variation induced solely by thermal expansion in the temperature interval 540-900 °C is shown for comparison. Reproduced with permission from Appl. Phys. Lett. 99, 122508. Copyright 2011, AIP Publishing LLC.

At the onset of detectability (at 540 °C), the mean Nb₃Sn crystallite size estimated from the Nb₃Sn (200) peak width is about 60 nm. During the 700 °C plateau the average crystallite size increases from 110 nm to about 180 nm. At the same time the Nb₃Sn volume increases by about 35% and the Nb₃Sn lattice parameter increases from 5.3140 to 5.3156 A. This indicates that the average Sn content increases by more than 1%, which in turn corresponds to a strong critical field $B_{c20}$ increase of 5 T [xxii]. At high magnetic field such a strong $B_{c2}$ increase outweighs the reduction of flux pinning force due to the simultaneous Nb₃Sn grains growth. Further increase in temperature and HT duration only slightly increases the Nb₃Sn volume but has a detrimental influence on the Nb₃Sn crystallite size, which results in limited $J_c$ when flux pinning has a dominating influence.

### G1.1.2.2.3 Nb₃Sn texture formation

Static texture analysis of bulk materials can be performed best by neutron diffraction measurements because of the relatively low neutron absorption. Synchrotron XRD in transmission geometry using an area detector can be very fast and is therefore better suited for monitoring texture formation *in situ* during processing HT [xxvii]. Diffraction images of different Nb₃Sn wires acquired with a Trixell Pixium 4700 twodimensional flatpanel digital detector are compared in Figure G1.1.2.6. Preferential crystallite orientation is revealed by intensity fluctuations along the Nb and Nb₃Sn diffraction rings [viii].

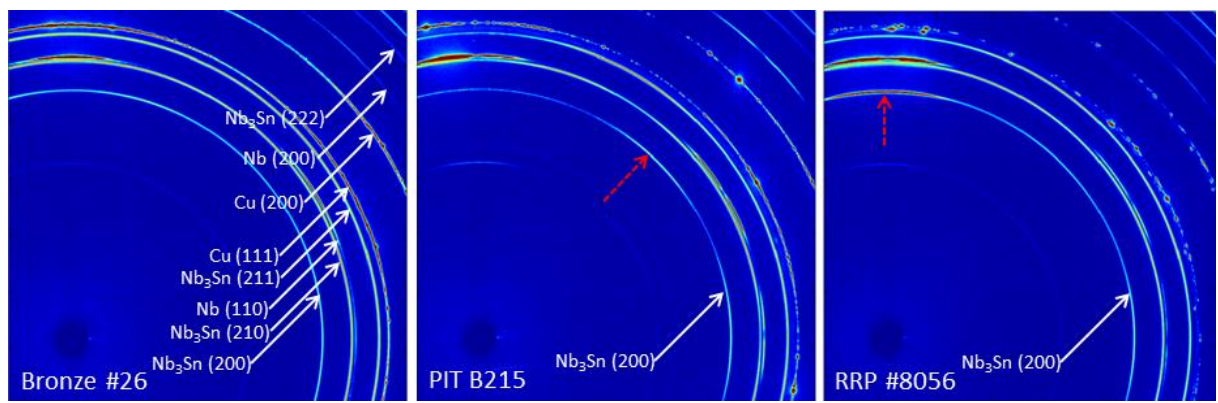

**Figure G1.1.2.6.** Diffraction pattern of a Bronze Route, PIT and RRP Nb₃Sn wire. The dashed arrows indicate the positions of intensity maxima in the Nb₃Sn (200) Debye rings. © IOP Publishing. Reproduced with permission. All rights reserved.

The homogeneous intensity along the Nb₃Sn rings of the BR wire shows that the BR process produces randomly oriented Nb₃Sn crystallites. In contrast, intensity maxima are seen in the Nb₃Sn (200) rings of the RRP and PIT type wires, but the intensity maxima are in different positions. Further texture analysis by Electron Backscatter Diffraction (EBSD) revealed a Nb₃Sn <110> texture in the PIT type wire, while in the RRP type wire Nb₃Sn grows with a <100> texture in the wire axis direction [xxviii]. EBSD also confirmed a strong <110> Nb texture parallel to the wire axis, as it is commonly observed in cold drawn body centered cubic (bcc) Nb.

### G1.1.2.2.4 Void growth mechanisms in Nb₃Sn superconductors

The presence of porosity in superconductors is often unavoidable, and the fabrication route can have a strong influence on the porosity volume and the distribution of voids that remains in the fully processed superconductor. Porosity generally reduces the useful superconductor volume in the composite, and in some cases it may degrade the irreversible strain limit of brittle superconductors. If the porosity is distributed inside the superconducting phase it can block the supercurrent.

The visualisation and quantitative description of the distribution of voids in the superconductor can help to better understand the porosity formation and redistribution mechanisms, and how porosity influences the superconducting properties. When the void shape and distribution are irregular, two-dimensional metallographic observations of void formation can be erratic and misleading. In contrast, X-ray micro-tomography (μ-CT) can provide non-destructively three-dimensional (3D) quantitative information about the porosity and particle size distribution.

At modern synchrotrons tomograms can be acquired in less than one minute, which enables time resolved *in situ* μ-CT studies of entire processes. Figure G1.1.2.7 shows a sequence of tomograms that were acquired *in situ* during the processing HT of a low Sn content Internal Tin Nb₃Sn wire [xxi] with a ramp rate of 60 °C/h, using the tomography furnace shown in Figure G1.1.2.2.(a). In order to obtain a 3D view of the porosity inside the wire, the strand materials have been transparently depictured in the image reconstructions.

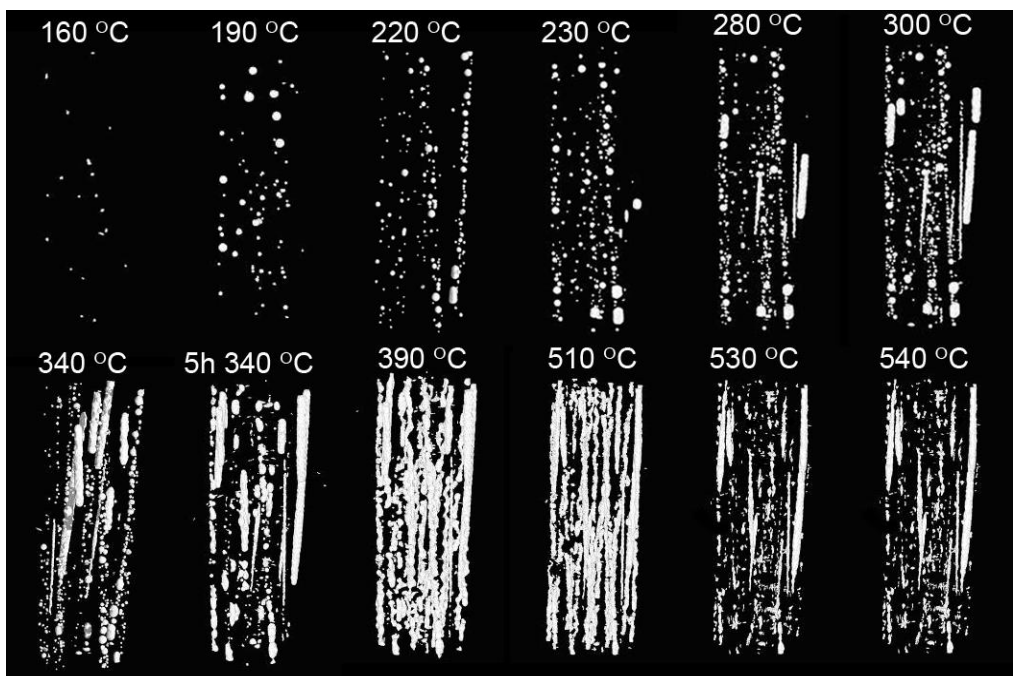

**Figure G1.1.2.7.** 3D view of the porosity inside an Internal Tin Nb₃Sn wire acquired *in situ* by synchrotron μ-CT at different temperatures. Reproduced with permission from Appl. Phys. Lett. 90, 132510. Copyright 2007, AIP Publishing LLC.

In Figure G1.1.2.8 the phase evolution during the HT, based on diffraction peak area measurements, is compared with the porosity volume evolution, which is determined from the simultaneously acquired tomograms (Figure G1.1.2.7).

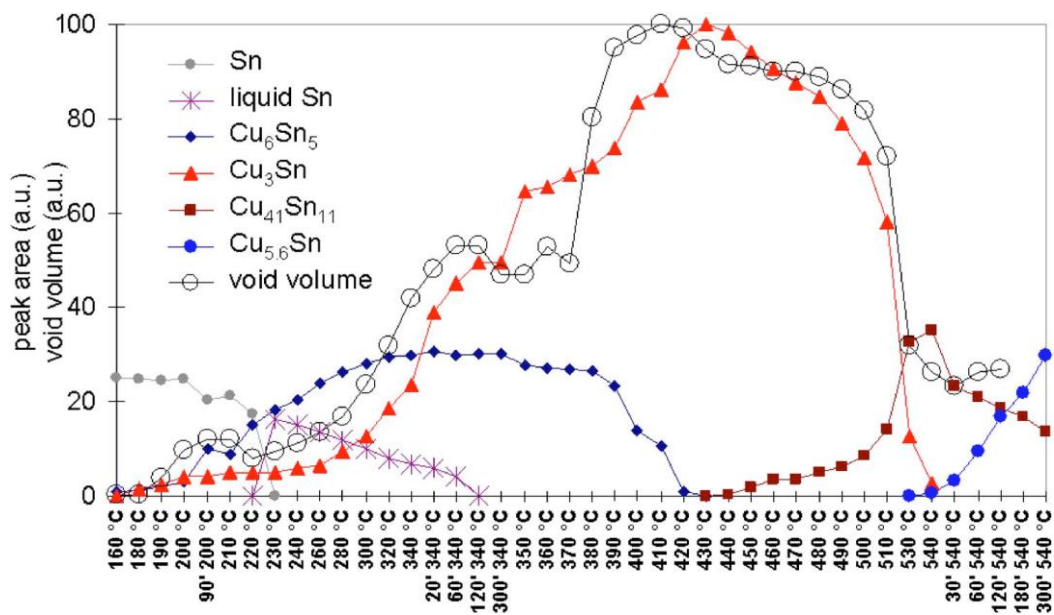

**Figure G1.1.2.8.** Evolution of prominent diffraction peak areas of all Sn containing phases, apart from α-bronze, that exist in the IT Nb₃Sn strand during the reaction HT up to 540 °C. Diffraction peak areas have been scaled such that the values correspond with the relative phase volume in the wire. The liquid Sn evolution is estimated from the amount of the detected phases. The total void volume is shown for comparison. Reproduced with permission from Appl. Phys. Lett. 90, 132510. Copyright 2007, AIP Publishing LLC.

The phase evolution during the HT of the low Sn content Internal Tin wire differs strongly from that of high Sn content PIT and RRP type wires (Figure G1.1.2.3). In particular the Nb containing phases $NbSn_2$, $CuNbSn$ and $Nb_6Sn_5$ are not formed in the low Sn content wire. The analysis of the simultaneously acquired µ-CT and XRD results allows to distinguish between different void formation mechanisms.

The growth of the globular voids up to a temperature of about 200 °C is driven by a gain in free energy through a reduction of the total void surface area when smaller voids present in the as-drawn wire agglomerate to larger globular voids. At 200 °C the maximum ratio of void volume to void surface area is obtained. At this temperature the total void volume corresponds to 2.5% of the pure Sn volume in the as-drawn wire. The correlation between void volume and $Cu_3Sn$ content, which is obvious in Figure G1.1.2.8, is due to the 4% higher density of $Cu_3Sn$ with respect to the Cu and Sn in their stoichiometric quantities.

## G1.1.2.3 Transformation HT of rapidly quenched $Nb_3Al$ precursor

The $Nb_3Al$ phase in superconducting wires is produced during a Rapid Heating Quenching and Transformation (RHQT) process [xxix]. During a HT at roughly 1900 °C a $Nb(Al)_{SS}$ solid solution is obtained, which can be retained during rapid quenching to ambient temperature. The Rapid Heating and Quenching (RHQ) stages are followed by a transformation HT with a peak temperature of typically 800 °C, during which fine grained $Nb_3Al$ with high Al content is formed from the $Nb(Al)_{SS}$ solid solution.

The phase evolution during this transformation HT can be studied *in situ* by high energy synchrotron X-ray diffraction [xxx]. The two-dimensional diffraction pattern acquired in transmission geometry with an area detector can be caked into sectors, in order to distinguish between reflections from the crystalline planes oriented both perpendicular and parallel to the wire drawing axis, which are in the following referred to as the axial and transverse directions, respectively.

The pattern presented in [xxx] have been caked into 128 sectors. The 222 filament $Nb_3Al$ precursor wire without Cu stabiliser that was studied has a partial interfilamentary Ta matrix. Since the strain free Ta and Nb lattice parameters at RT differ by about 0.03 % only, these phases could not be distinguished by their lattice spacing. Therefore, in the following Nb diffraction peak refers to both overlapping peaks of Nb and Ta.

The axial and transverse Nb (110) diffraction peaks of the RHQ wire are presented in Figure G1.1.2.9. The axial Nb (110) peak is about 8 times more intense than the transverse peak, which shows that the Nb (and/or Ta) texture, which is developed during the cold drawing of bcc metals, is partially retained during the RHQ process. The axial and transverse Nb peaks exhibit two maxima, which are characteristic for pure Nb and Ta (larger d-spacing) and $Nb(Al)_{SS}$ supersaturated solid solution (with roughly 1% smaller d-spacing).

The evolution of the Nb (110), $Nb_3Al$ (200) and $Nb_3Al$ (211) diffraction peak shape and intensity during the RHQ $Nb_3Al$ precursor wire transformation HT with a ramp rate of 800 °C/h and a final 800 °C plateau lasting 30 minutes can be seen in Figure 5 of reference [xxx]. The Nb (110) peak shape change is caused by the vanishing of the $Nb(Al)_{SS}$ peak component, upon formation of $Nb_3Al$. When heating with a ramp rate of 800 °C/h $Nb_3Al$ (200) and $Nb_3Al$ (211) peaks are detected at about 780 °C. When heating with a ramp rate of 160 °C/h the transformation from a Nb(Al)$_{SS}$ supersaturated solid solution into Nb$_3$Al occurs at roughly 60 °C lower temperature than during the 800 °C/h HT [xxx].

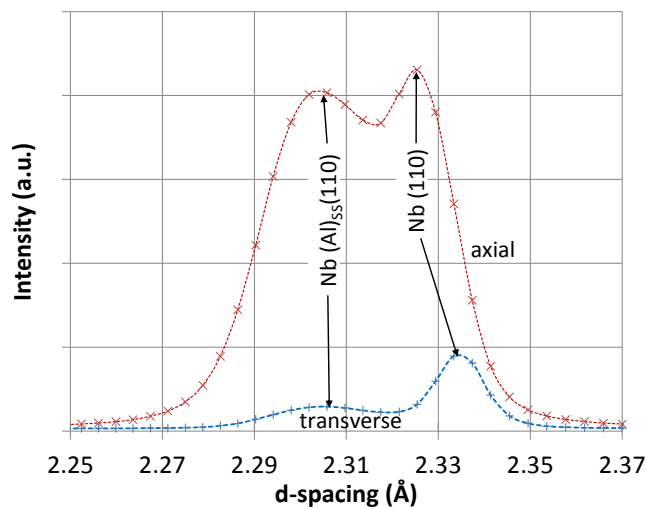

**Figure G1.1.2.9.** Axial and transverse Nb(110) diffraction peak, consisting of two components characteristic for pure Nb and for Nb(Al)$_{SS}$.

## G1.1.2.4 Bi-2212 wire melt processing

### G1.1.2.4.1 Phase evolution during Bi-2212 wire melt processing

In order to form well connected and textured Bi-2212 filaments, the Bi-2212 precursor particles in the as-drawn Bi-2212 PIT wire need to be melted when the wire is at its final size and shape [xxxi]. During the melt processing HT an external oxygen supply through the oxygen permeable Ag wire matrix is needed in order to re-form Bi-2212 out of the melt. The phase evolution during the melt processing HT can be studied *in situ* by high energy synchrotron XRD measurements. Oxygen can be supplied conveniently in a flow of air at ambient pressure, using the X-ray transparent furnace shown in Figure G1.1.2.2(a).

The sequence of diffraction pattern acquired during the melt processing of a state-of-the-art Bi-2212 PIT wire in air (oxygen partial pressure pO$_2$=0.21 bar) is presented in Figure G1.1.2.10 [xxxii]. An initial Bi-2212 diffraction peak growth with increasing temperature is observed, which is attributed to crystallization of Bi-2212 that was amorphized during the wire drawing process. The main impurity phase Bi-2201 is first detected when the temperature exceeds approximately 200 °C and a maximum amount of Bi-2201 is detected at about 500 °C. Bi-2201 decomposes completely at 850 °C, and reforms again upon cooling at approximately 850 °C. The diffraction peaks that occur upon Bi-2212 melting around 880 °C in a pO$_2$=0.21 bar process gas have been tentatively identified as Cu-free phase Bi$_2$(Sr$_{4-y}$Ca$_y$)O$_7$.

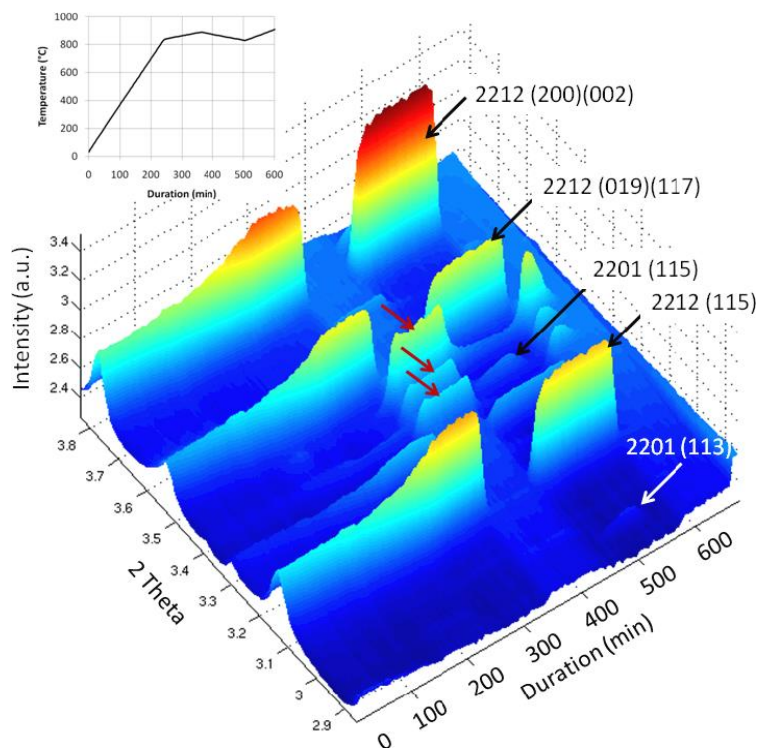

**Figure G1.1.2.10.** Sequence of XRD patterns acquired during Bi-2212 wire HT in ambient air. The diffraction peaks which are labelled with arrows have been tentatively identified as the Cu-free phase Bi₂(Sr₄₋ᵧCaᵧ)O₇). © IOP Publishing. Reproduced with permission. All rights reserved.

Overpressure (OP) processing at pressures of up to 100 bar is a key for achieving homogeneous high critical currents in long lengths of Bi-2212 wires [xxxiii]. OP processing also enables varying the oxygen partial pressure in a wide range and it is of interest to verify how $pO_2$ influences the phase sequence and the Bi-2212 precursor melting and recrystallization behaviours.

In order to study the influence of $pO_2$ on the Bi-2212 phase stability inside the Bi-2212/Ag wire by *in situ* high energy synchrotron XRD measurements, the high pressure cell and capillary furnace shown in Figure G1.1.2.2.(b,c) have been used. This furnace allows to explore $pO_2$ above ambient pressure, with total process gas pressures up to 200 bar. Another advantage of this furnace is that 5 cm-long wire samples with closed ends identical to the samples typically used for Bi-2212 critical current measurements can be studied.

The diffraction pattern acquired *in situ* during HTs at different $pO_2$ show that increasing $pO_2$ reduces the Bi-2212 stability [vi]. At $pO_2$=1.5 bar Bi-2212 decomposes partly prior to melting, and the precursor decomposition temperature is about 20 °C lower than it is at $pO_2$=1.05 bar. At $pO_2$=5 bar the Bi-2212 precursor particles in the state-of-the art Bi-2212 multifilament wire decomposes completely in the solid state.

### G1.1.2.4.2 Void formation and redistribution during Bi-2212 wire melt processing

Porosity and second phase particles formed during melt processing are considered to be the main current limiting defects in Bi-2212 wires. It is therefore of great interest to visualise and to quantify the porosity and second phase distribution during the different processing steps.

The potential of µ-CT to visualise these features inside a superconducting wire depends equally on the spatial and density resolution of the µ-CT experiment. The calculated linear absorption coefficients of 70 keV photons in the main wire constituents Ag and Bi-2212, and the main impurity phase Bi-2201, are $\mu_{Ag}$=40 cm$^{-1}$, $\mu_{Bi-2212}$=15 cm$^{-1}$ and $\mu_{Bi-2201}$=18 cm$^{-1}$, respectively. Because of the different X-ray attenuation in Ag, Bi-2212 and porosity, high energy synchrotron µ-CT is particularly well suited to monitor Bi-2212 microstructure changes and the porosity formation and redistribution inside Bi-2212/Ag wires [xxxii]. On the other hand, because of the relatively small difference of the X-ray attenuation in Bi-2212 and Bi-2201, these phases cannot be distinguished in the X-ray absorption tomograms.

The void redistribution during the melt processing of a 37x17 filament Bi-2212 wire at ambient pressure can be followed in the longitudinal µ-CT cross sections shown in Figure G1.1.2.11, which have been acquired *in situ* at different temperatures. In order to show a more detailed view of the voids the images have been cropped from the longitudinal cross sections showing the entire wire cross section. The black areas represent voids, the bright grey areas the strongly absorbing Ag matrix, and the dark-grey areas are Bi-2212 with a small amount of Bi-2201.

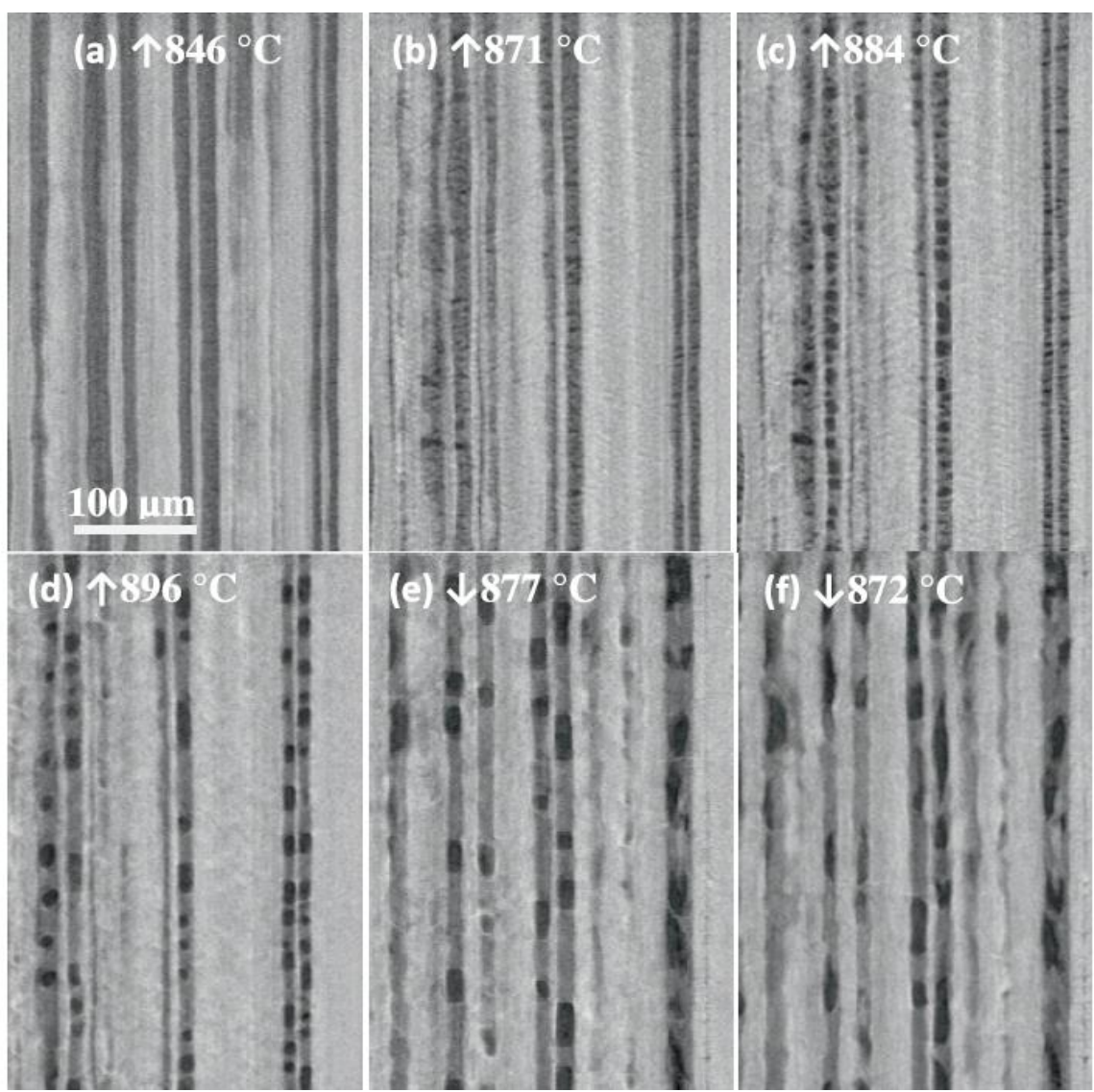

**Figure G1.1.2.11.** Detailed view of Bi-2212 wire longitudinal tomographic cross sections acquired *in situ* at different temperatures during HT to T$_{max}$=915 °C in air. A time lapse movie

showing the changes occurring over the whole heating and cooling cycle is available ( https://edms.cern.ch/document/1153082/1 ). © IOP Publishing. Reproduced with permission. All rights reserved.

Filament microstructure changes can be first observed at about 850 °C when the Bi-2201 impurity phase decomposes (as seen in the simultaneously acquired XRD pattern). At this temperature the finely divided porosity, which is in the as-drawn wire uniformly distributed between the precursor particles, coalesces into lens-shaped defects. On Bi-2212 melting the lens-shaped voids grow to bubbles of a filament diameter.

Upon cooling nucleation of Bi-2212 is first observed in the tomogram acquired at 877 °C at the filament periphery. The Bi-2212 formed upon cooling partly bridges the void space, but bubbles remain and cause an obstacle to the current flow in the Bi-2212 wires that are melt processed at ambient pressure [xxxiv].

The importance of complete Bi-2212 precursor melting is obvious when comparing the longitudinal µ-CT cross section acquired at the end of a processing HT to $T_{max}$=915 °C, during which Bi-2212 was completely melted (G.1.1.2.12(a)), and to $T_{max}$=875 °C, in which only a fraction of the Bi-2212 powder was melted (G.1.1.2.12(b)). The tomograms show clearly that after the $T_{max}$=875 °C HT the filaments remain interrupted by a regular array of lens shaped voids, and that filament connectivity is only achieved after the porosity rearrangement that occurs during complete Bi-2212 melting and recrystallization.

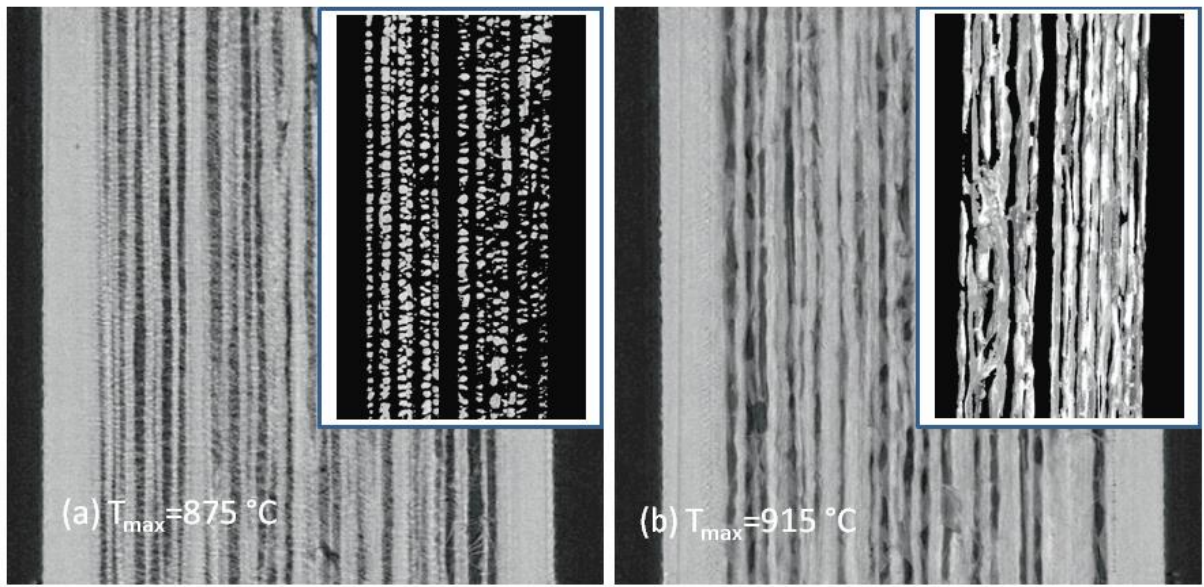

**Figure G1.1.2.12.** Tomographic cross sections of the Bi-2212 wire acquired after *in situ* HT to (a) $T_{max}$=875 °C and (b) $T_{max}$=915 °C. 3D reconstructed images of selected filaments are shown in the insets. © IOP Publishing. Reproduced with permission. All rights reserved.

The overall porosity volume in Bi-2212 wires that are short enough to allow relief of internal pressure through the open ends does not strongly change, because the Bi-2212 processing does not involve phase transformations associated with important density variations, as it is for instance the case in $Nb_3Sn$ conductors [xii]. In long Bi-2212 wires and in wires with closed ends, additional porosity is formed during processing at ambient pressure when internal gas pressure leads to creep of the Ag matrix [xxxv]. OP pressing strongly reduces the porosity volume that is present in the as-drawn wire [xxxiii].

## G1.1.2.5 Outlook

Today the time resolved combined XRD and μ-CT experiments for *in situ* studies of superconductors that are described above can be routinely performed at advanced high energy synchrotron beamlines.

The continuously improving brilliance of synchrotron sources and new efficient X-ray focusing optics make it possible to use nanometer scale X-ray beams, enabling new non-destructive *in situ* experiments on length scales that so far were only accessible to destructive techniques [xxxvi].

Grain size and grain orientation have a dominant influence on the performance of most superconductors, and studies of the thermal growth of grains and the grain orientation evolution are examples were future superconductor research can profit from new synchrotron experiments with X-ray nanobeams. Such studies can be performed in two dimensions, averaging over the sample depth that is penetrated by the X-ray beam. When applying tomographic methods (e.g. XRD-tomography [xxxvii]) spatially resolved *in situ* studies of phase composition, crystallite size distribution and texture become possible.

### Acknowledgments

All XRD and μ-CT experiments presented here have been performed at the ESRF ID15 beamline. We are grateful to Julian Kadar for the $Nb_3Sn$ diffraction peak analysis of Figure G1.1.2.4.